\documentclass[preprintnumbers,superscriptaddress,nofootinbib,aps,prd,twocolumn ]{revtex4}

\pdfoutput=1

\usepackage{multirow,array}
\usepackage{amsmath,amssymb}
\usepackage{graphicx}
\usepackage{color}
\usepackage{url}
\usepackage{slashed}
\usepackage{feynmp}
\usepackage[alsoload=hep]{siunitx}
\usepackage[pdfborder={0 0 0}]{hyperref}
\hypersetup{
  pdfauthor={Dorival Goncalves, Kazuki Sakurai, Michihisa Takeuchi},
  pdftitle={Tagging a mono-top signature in Natural SUSY},
  pdfkeywords={QCD, NLO, Higgs Physics}
}
\def\beq{\begin{equation}}
\def\eeq{\end{equation}}
\def\beqn{\begin{eqnarray}}
\def\eeqn{\end{eqnarray}}

\begin{document}

\title{  Tagging a mono-top signature in Natural SUSY}

\author{ Dorival Gon\c{c}alves}
\affiliation{Institute for Particle Physics Phenomenology, Department of Physics, Durham University, UK}
\affiliation{PITT-PACC,  Department  of  Physics  and Astronomy,  University  of  Pittsburgh,  USA \\[0.1cm]}
\author{ Kazuki Sakurai}
\affiliation{Institute for Particle Physics Phenomenology, Department of Physics, Durham University, UK}
\affiliation{Institute of Theoretical Physics, Faculty of Physics, University of Warsaw,  Poland}
\author{ Michihisa Takeuchi}
\affiliation{Kavli IPMU (WPI), UTIAS, University of Tokyo, Japan}




\begin{abstract}
  \noindent
We study the feasibility of probing  a region of Natural Supersymmetry  where 
the stop and higgsino masses are compressed.
Although this region is most effectively searched for in the mono-jet channel,
this signature is present in many other non-supersymmetric frameworks.
Therefore, another channel that carries orthogonal information is required to confirm the
existence of the light stop and higgsinos.
We show that a supersymmetric version of the $t \bar t H$ process, 
$pp \to t \tilde t_1 \tilde \chi^0_{1(2)}$,
can have observably large rate when both the stop and higgsinos are significantly light,
and it leads to a distinctive mono-top signature in the compressed mass region.
We demonstrate that the hadronic channel of the mono-top signature 
can effectively discriminate the signal from backgrounds by tagging a hadronic top-jet.
We show that the hadronic channel of mono-top signature offers a significant improvement 
over the leptonic channel and the sensitivity reaches $m_{\tilde t_1} \simeq 420$ GeV
at the 13 TeV LHC with 3\,ab$^{-1}$ luminosity.

\end{abstract}
\preprint{IPPP/16/96,~IPMU16-0155,~PITT-PACC-1611}

\maketitle


\section{Introduction}
\label{sec:intro}

The experiment at CERN's Large Hadron Collider (LHC) has discovered a Higgs-like boson~\cite{Chatrchyan:2012xdj,Aad:2012tfa}, 
yet  no sign of new physics beyond the Standard Model has been seen~\cite{Corbett:2015ksa}.
The gauge hierarchy problem of the Standard Model becomes more compelling than ever before.  
The most promising solution to the gauge hierarchy problem is low energy supersymmetry (SUSY),
where the radiative correction to the Higgs mass squared parameter from Standard Model particles
is cancelled by the contribution from their superpartners, and the electroweak scale is stabilised if sparticles are not significantly heavier than {\cal O}(100) GeV.
The null result in SUSY searches at the LHC pushes the mass limit for sparticles 
and creates a tension between the two scales: the naturally expected electroweak scale and the observed one.
One way to relax this tension is to arrange the mass spectra such that
all SUSY particles are safely beyond the current mass limit
but keep the scalar top-quark (stop) and the higgsinos as light as possible. 
This solution is dubbed as {\it Natural SUSY} and has been intensively studied~\cite{Kitano:2006gv,
Papucci:2011wy,
Hall:2011aa,
Desai:2011th,
Ishiwata:2011ab,
Sakurai:2011pt,
Kim:2009nq,
Wymant:2012zp,
Baer:2012up,
Randall:2012dm,
Cao:2012rz,
Asano:2012sv,
Baer:2012uy,
Evans:2013jna,
Hardy:2013ywa,
Kribs:2013lua,
Bhattacherjee:2013gr,
Rolbiecki:2013fia,
Curtin:2014zua, 
Kim:2014eva,
Papucci:2014rja,
Casas:2014eca,
Katz:2014mba,
Heidenreich:2014jpa,
Mustafayev:2014lqa,
Brummer:2014yua,
Cohen:2015ala,
Hikasa:2015lma,
Barducci:2015ffa,
Drees:2015aeo,
Mitzka:2016lum,
Graesser:2012qy,
Basirnia:2016szw,
Kim:2016rsd,
Han:2016xet,
Kouda:2016ifp,
Han:2016hgr}.

The light stop scenario has also attracted a lot of attention in the experimental community 
and many analyses have been dedicated to the light stop search.
One of the most challenging parameter region is so-called {\it compressed region}, where
the lighter stop, $\tilde t_1$, is only slightly heavier than the lightest neutralino, $\tilde \chi^0_1$, which is assumed to be the lightest SUSY particle (LSP) and stable.    
In this region, the decay products of the stop become very soft and are not reconstructed in the detector.
The total missing energy becomes also very small due to the back-to-back kinematics of the stop pair.

\begin{figure*}[t!]
 \begin{center}
          \includegraphics[scale=0.38]{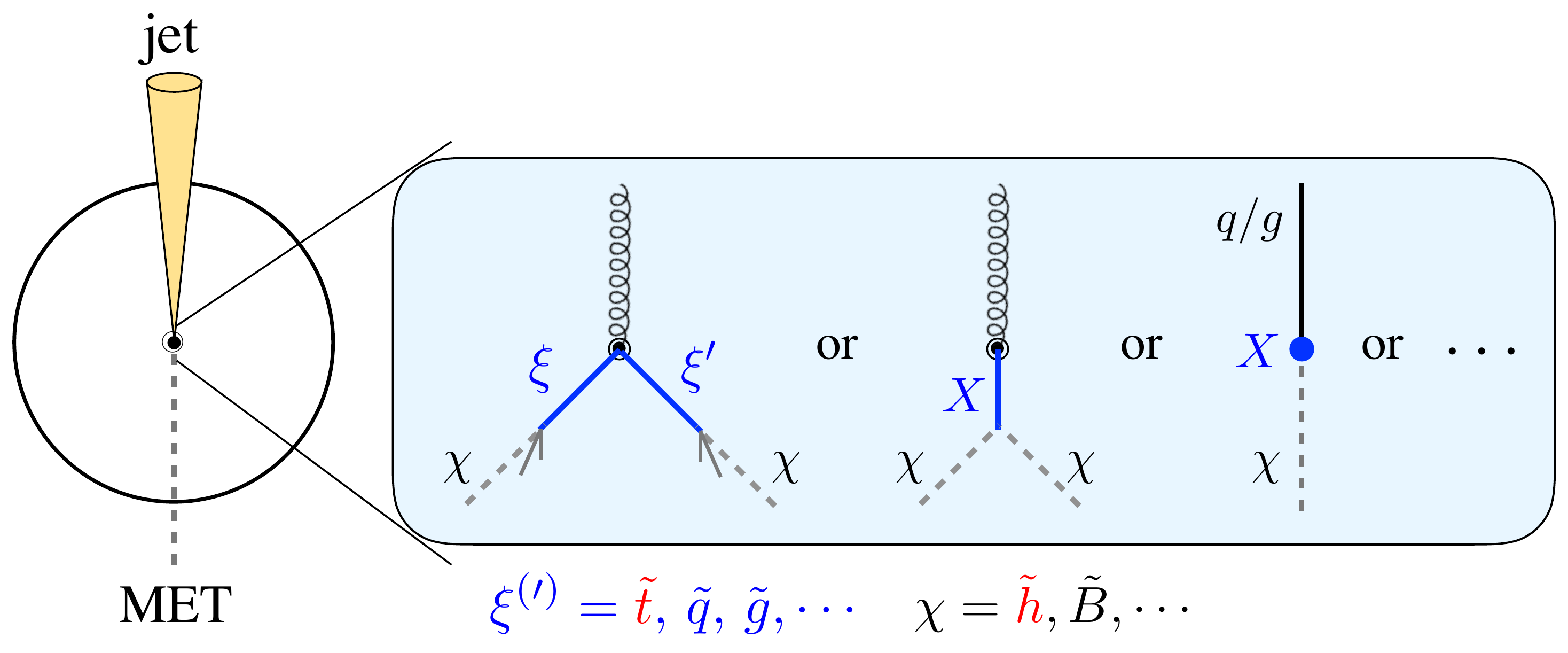}          
          \hspace{10mm}
          \includegraphics[scale=0.28]{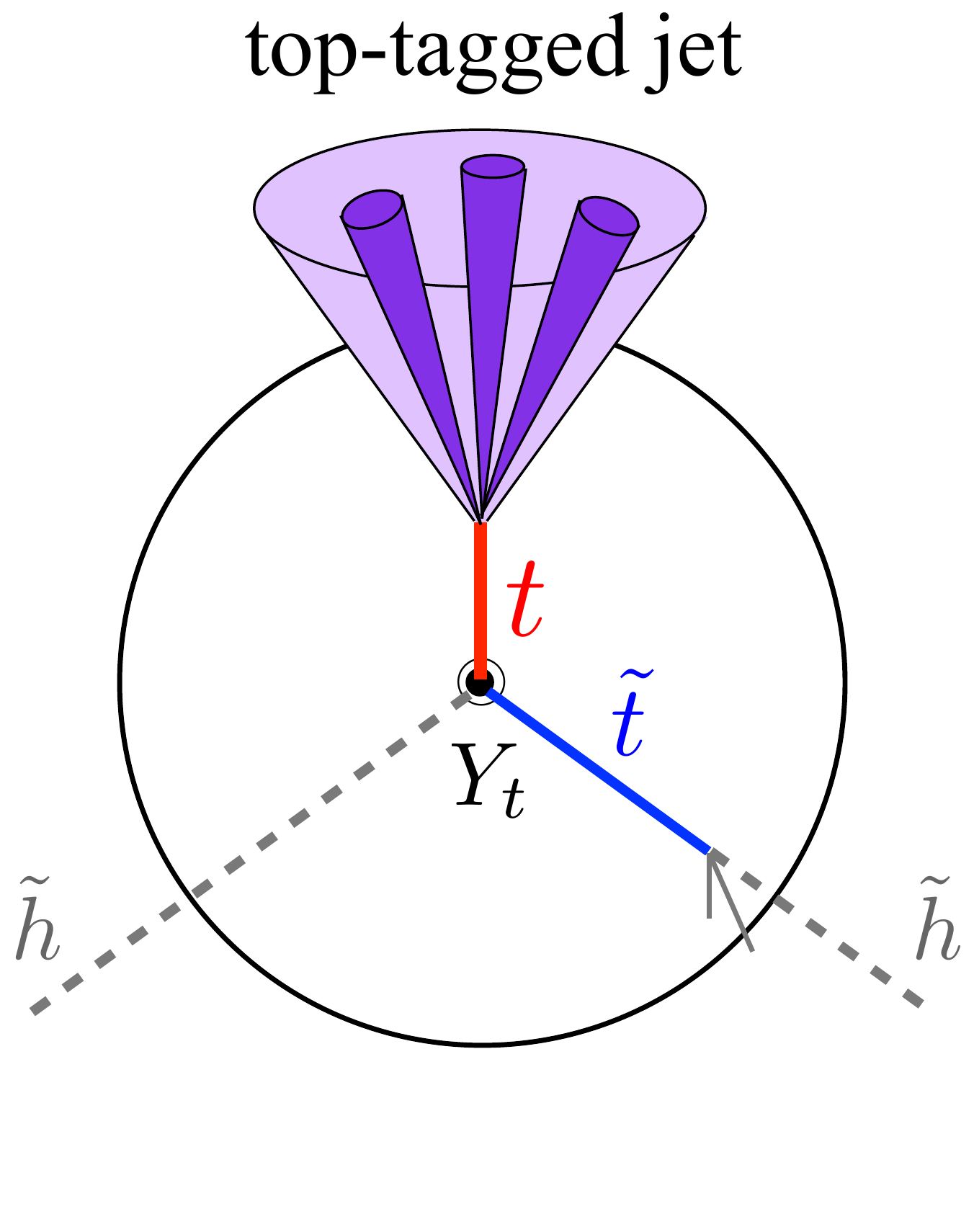}                    
 \caption{Mono-jet signature (left) and hadronic mono-top signature (right).}
 \label{fig:evdisp2}
\end{center}
\end{figure*}

The compressed stop-neutralino region is intensively searched for by looking at mono-jet signature~\cite{Aaboud:2016tnv, Khachatryan:2016pxa}
where the stop pair system  is boosted recoiling against hard QCD initial state radiation, creating a large missing energy. 
Although this search channel is very powerful in terms of discovery, there is an important drawback.
Its final state is characterised by large missing energy associated with  high $p_T$ jet(s), and none of the high $p_T$ objects comes 
directly from stops. Indeed, as illustrated in the left panel of Fig.~\ref{fig:evdisp2}, the produced particles $\xi$ and $\xi^\prime$ are not necessarily stops but may be anything as long as they convert into the missing particle $\chi$, producing only soft particles that cannot be reconstructed in the detector. The same final state can also be realised by a single production of $X$ accompanied by hard QCD radiation followed by
an invisible decay $X \to \chi \chi$ or a resonant production of $X$ followed by $X \to q(g) + \chi$.
The list goes on. Finding a mono-jet signature thus by no means indicates the existence of a light stop nor the solution to the hierarchy 
problem~\cite{Khachatryan:2014rra,Harris:2014hga,Buckley:2014fba}.

In Ref.~\cite{Goncalves:2016tft}, we have pointed out that in addition to the mono-jet channel, the light stop and higgsinos,
if present in nature, must generate another phenomenologically attractive channel, namely $pp \to t \tilde t_1 \tilde \chi^0_{i}$ $(i=1,2)$\footnote{We do not explicitly distinguish the particle and the anti-particle in writing $pp \to t \tilde t_1 \tilde \chi^0_{i}$.
The baryon and flavour numbers are however always conserved in this process.
}.
The relation between $pp \to \tilde t_1 \tilde t_1^*$ and ${pp \to t \tilde t_1 \tilde \chi^0_{i}}$
is analogous to $pp \to t \bar t$ and $pp \to t \bar t H$ in the Standard Model \cite{Aad:2015iha,
Aad:2014lma,
Khachatryan:2014qaa,
Degrande:2012gr,
Ellis:2013yxa,
Nishiwaki:2013cma,
Santos:2015dja,
Li:2015kaa,
Moretti:2015vaa,
Buckley:2015vsa},
except that the $pp \to t \tilde t_1 \tilde \chi^0_{i}$ process leads to a prominent mono-top signature~\cite{Aad:2014wza, Khachatryan:2014uma, Fuks:2014lva, Davoudiasl:2011fj, Kamenik:2011nb, 
Andrea:2011ws, Alvarez:2013jqa, Agram:2013wda, Kling:2012up,Boucheneb:2014wza}.
In Fig.~\ref{fig:evdisp2} right panel, we depict the mono-top signature in the hadronic final state.
In the compressed region, the decay of $\tilde t_1$ is not resolvable in the detector and both the $\tilde t_1$ and $\tilde \chi_1^0$ 
contribute to the missing energy, leaving the top-quark alone in the final state as a visible object.
Importantly, the event rate of this process can be observably large only if the neutralino is dominantly composed of higgsinos.
Therefore, the observation of this process is a strong indication for the existence of both the light stop and the higgsinos.

The leptonic final state of $pp \to t \tilde t_1 \tilde \chi^0_{i}$ has been studied in \cite{Goncalves:2016tft}.
The advantage of the leptonic channel is the ability of probing the left-right mixing of the $\tilde t_1$ by looking at angular distributions of the charged lepton and the $b$-jet originated from the top-quark decay \cite{Goncalves:2016tft}.
On the other hand, the event rate of this channel is limited due to the small top leptonic branching ratio 
and a partial cancellation in the missing energy between the neutralinos and the neutrino from the top-quark decay,
as we will discuss later in detail.

In this paper we study the hadronic channel of the $pp \to t \tilde t_1 \tilde \chi^0_{i}$ process, where an obvious advantage is the large hadronic branching ratio.
Unlike the leptonic channel, reconstructing hadronic top is non-trivial but crucial to discriminate the signal from the background.
We observe that in order to reduce the background, we necessarily require large missing energy, which forces the top-quark to be in a boosted regime.
In this regime, the hadronic top-quark can be reconstructed as a fat jet with a certain substructure in it, as depicted in the right 
panel of Fig.~\ref{fig:evdisp2}. In order to systematically ``tag'' the top-jets we use the \textsc{HepTopTagger}~\cite{Plehn:2009rk,Plehn:2010st} 
in our analysis. We find a significant improvement in the sensitivity over the leptonic channel of this process.
The paper is organised as follows.  In the next section we describe our analysis in detail and demonstrate
the top-jet tagging works well in conjunction with the large missing energy requirement.  
In section \ref{sec:results} we present our results. Finally, a summary of our key findings is presented in section~\ref{sec:conclusion}.

\section{Analysis}
\label{sec:analysis}

\begin{figure}[t!]
\centering
 \includegraphics[scale=0.55]{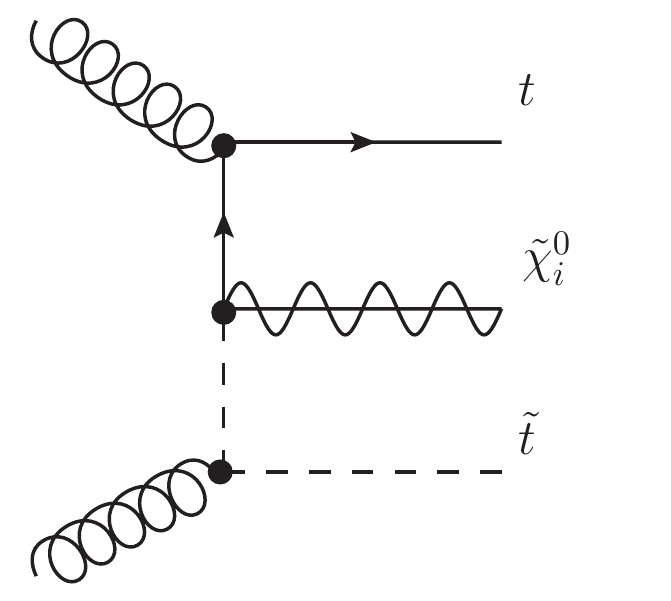}   
 \hspace{0.1cm}       
  \includegraphics[scale=0.56]{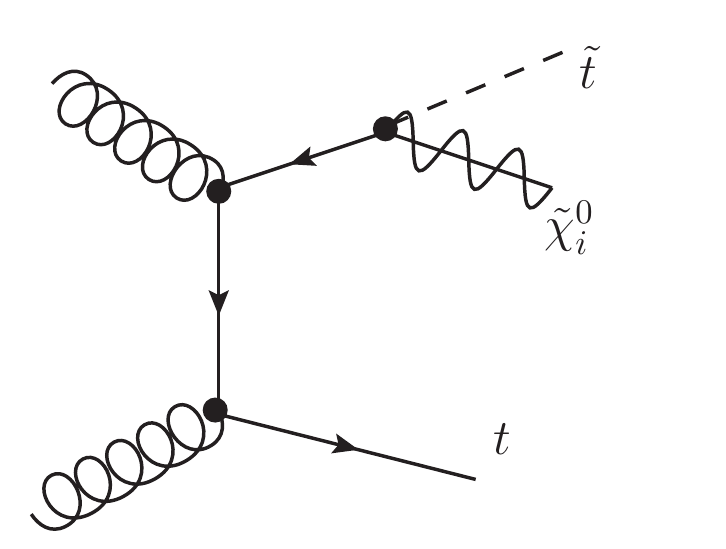}          
 \caption{Representative Feynmann diagrams for  ${pp \to t \tilde{t}_1\tilde{\chi}^0_{i}}$ $(i=1,2)$. }
 \label{fig:diagrams}
\end{figure}

We study the hadronic mono-top signature from the  $pp\rightarrow t \tilde{t}_1\tilde{\chi}_{1(2)}^0$ channel 
in the compressed  stop-higgsino mass region: $m_{\tilde{t}_1}<m_{\tilde{\chi}_{1(2)}^0}+m_W$. In Fig.~\ref{fig:diagrams}, we display a 
representative set of Feynman diagrams for this process. Unlike the mono-jet signature that exploits hard initial state radiation, 
our signal events possess large missing energies recoiling against a single boosted hadronic top.  
This channel therefore provides further information on the new physics interaction between the neutralino and stop sectors.  
We focus on the  Natural SUSY scenario where $\tilde{\chi}_1^0$ and $\tilde{\chi}_2^0$ are higgsino-like and almost mass degenerate.
In this case,
$t \tilde{t}_1\tilde{\chi}_{1}^0$ and $t \tilde{t}_1\tilde{\chi}_{2}^0$
processes contribute to the signal with almost equal rates. 
We also assume that the lighter chargino, $\tilde{\chi}_{1}^\pm$, is higgsino-like 
and almost mass degenerate with $\tilde{\chi}_1^0$.
The stops decay into $b$ and $\tilde{\chi}_1^\pm$ with 100\% branching ratio in our set up.
The major backgrounds for this search are $\bar{t}t$, $tZ$, $tW$ and $Z$+jets. 

The signal sample $pp \to t \tilde{t}_1\tilde{\chi}_{i}^0$ $(i=1,2)$  is generated with \textsc{MadGraph5+Pythia6}~\cite{mg5,pythia} and  
a flat Next-to-Leading Order $K$-factor of 1.5 is applied to rescale the leading-order cross section~\cite{Beenakker:1997ut,Beenakker:2010nq,Beenakker:2011fu,Binoth:2011xi,Goncalves:2014axa}. 
The backgrounds $\bar{t}t$ and $Z$+jets are produced with \textsc{Alpgen+Pythia6}~\cite{Mangano:2002ea}, merged up to one and three extra jets, respectively, with the MLM matching scheme.  
The $tZ$ and $tW$ backgrounds are generated with \textsc{SHERPA}~\cite{Gleisberg:2008ta}. 
For the $t\bar{t}$ background we normalize the sample to the NLLO+NLL cross section of 831\,pb~\cite{tt_NNLO}.  All signal 
and background samples include hadronisation and underlying event effects. The detector effects are simulated using the \textsc{Delphes3} package~\cite{deFavereau:2013fsa}.  

\begin{figure}[t!]
    \begin{center}
     \includegraphics[scale=0.32]{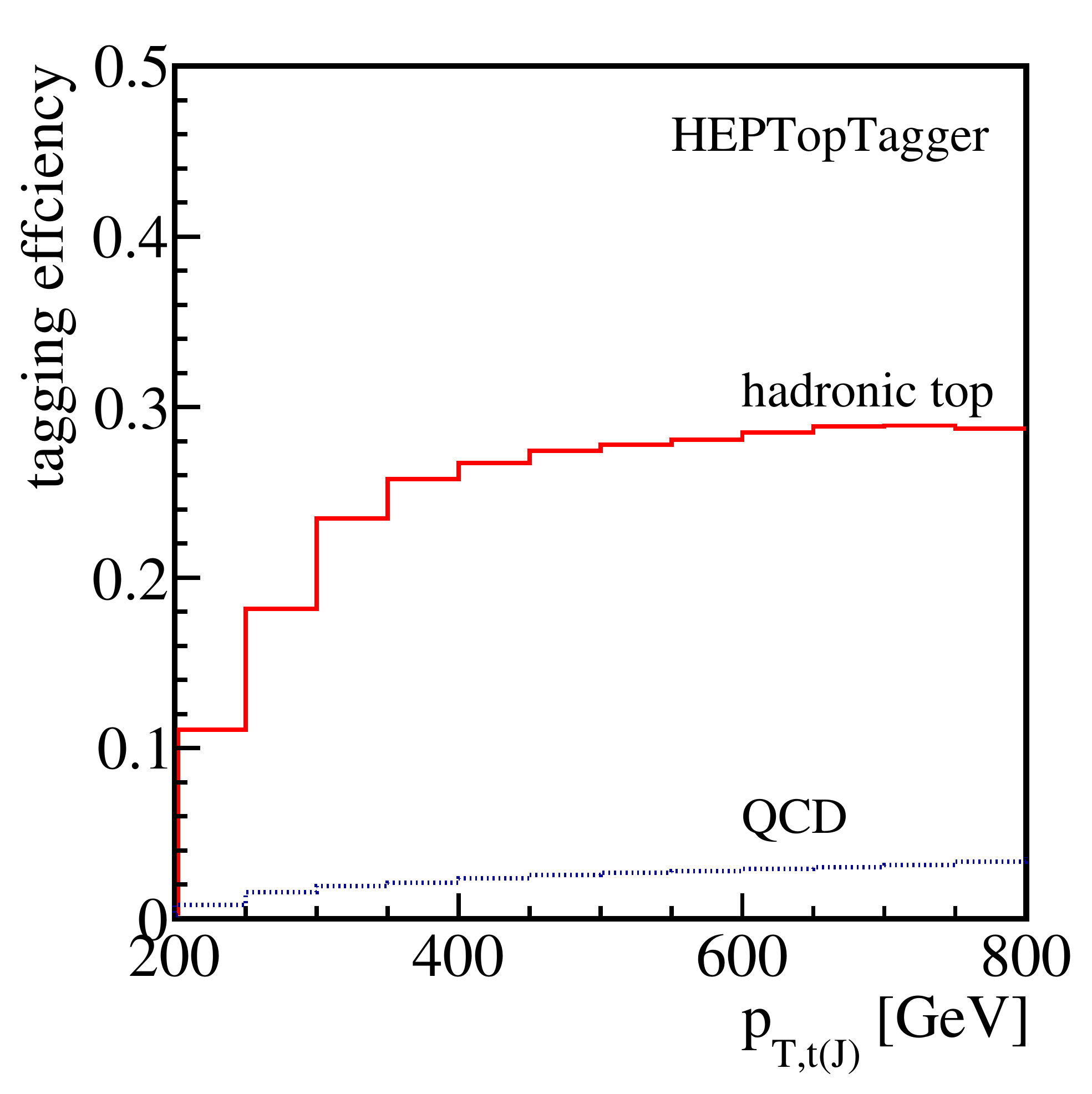}
      \caption{Top tagging efficiency as a function of the top-quark $p_{T,t}$ derived from the
    \textsc{HepTopTagger} algorithm. The dotted line shows the mistag rate as a function of the fat jet $p_{T,J}$ 
    originated from QCD jets in the $Z$+jets sample.}
    \label{fig:toptag}
    \end{center}
\end{figure}

\begin{figure}[t!]
    \begin{center}
          \includegraphics[scale=0.32]{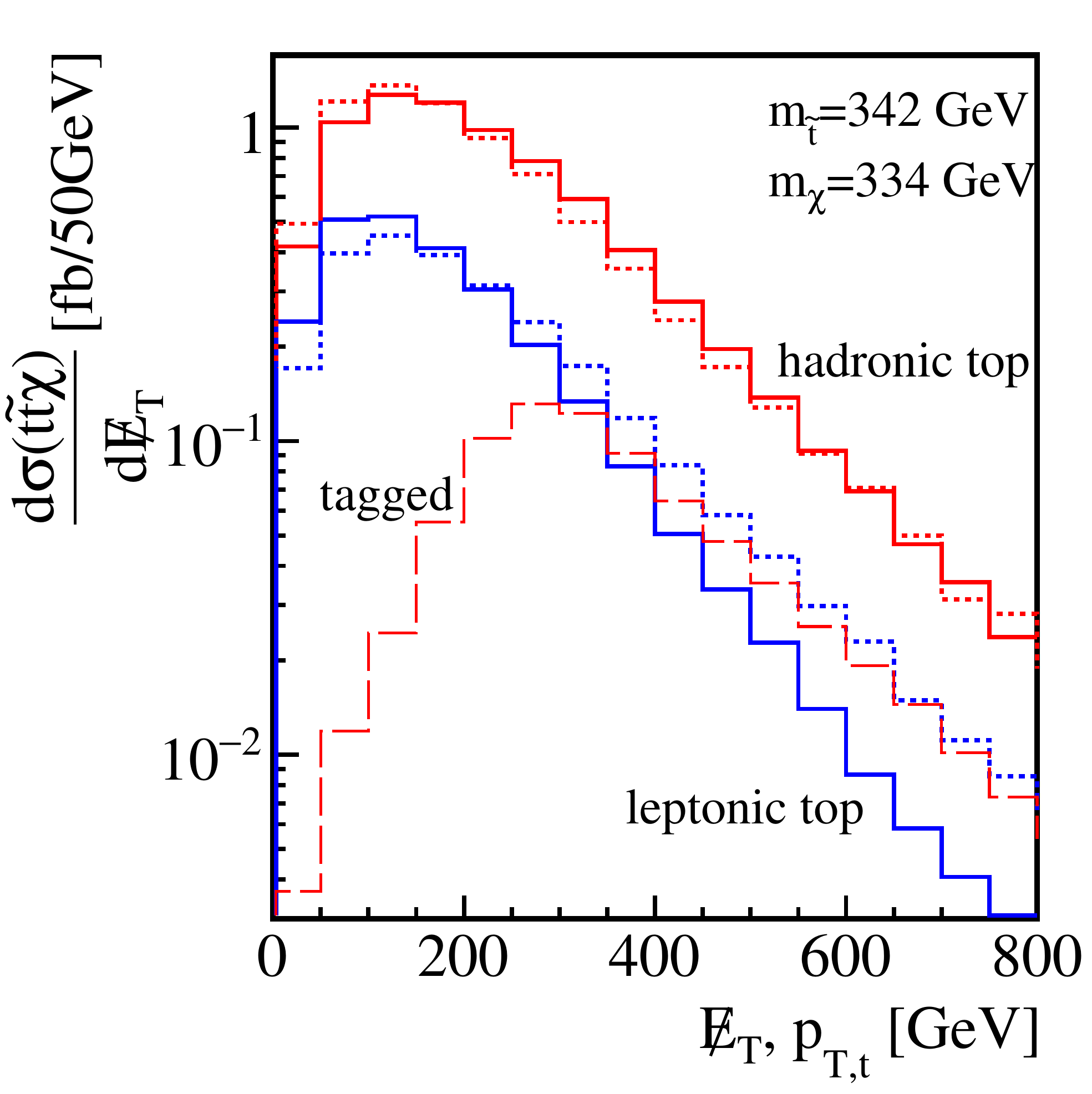}    
    \caption{$\slashed{E}_T$ distribution for leptonic  (blue) and hadronic (red) top decays for 
    $t\tilde{t}_1\tilde{\chi}_{1(2)}^{0}$ in the benchmark point of $(m_{\tilde t_1},m_{\tilde{\chi}_1^{0}})=$ (342~GeV, 334~GeV).  The dotted lines are the respective top $p_{T,t}$ distributions and the dashed line shows only the tagged contribution.
  }
    \label{fig:concept}
    \end{center}
\end{figure}

\begin{figure}[t!]
    \begin{center}
          \includegraphics[scale=0.32]{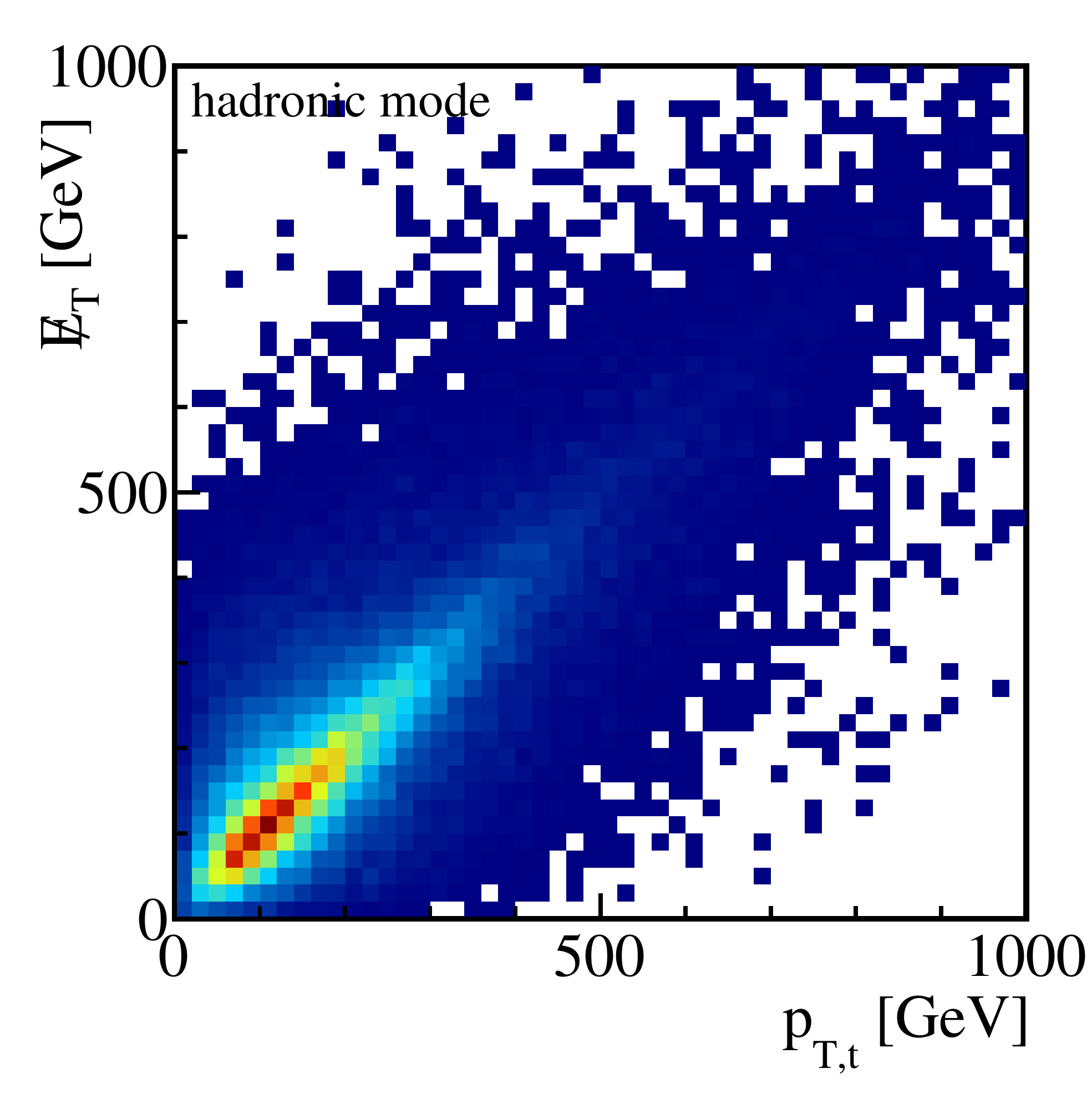}     
          \includegraphics[scale=0.32]{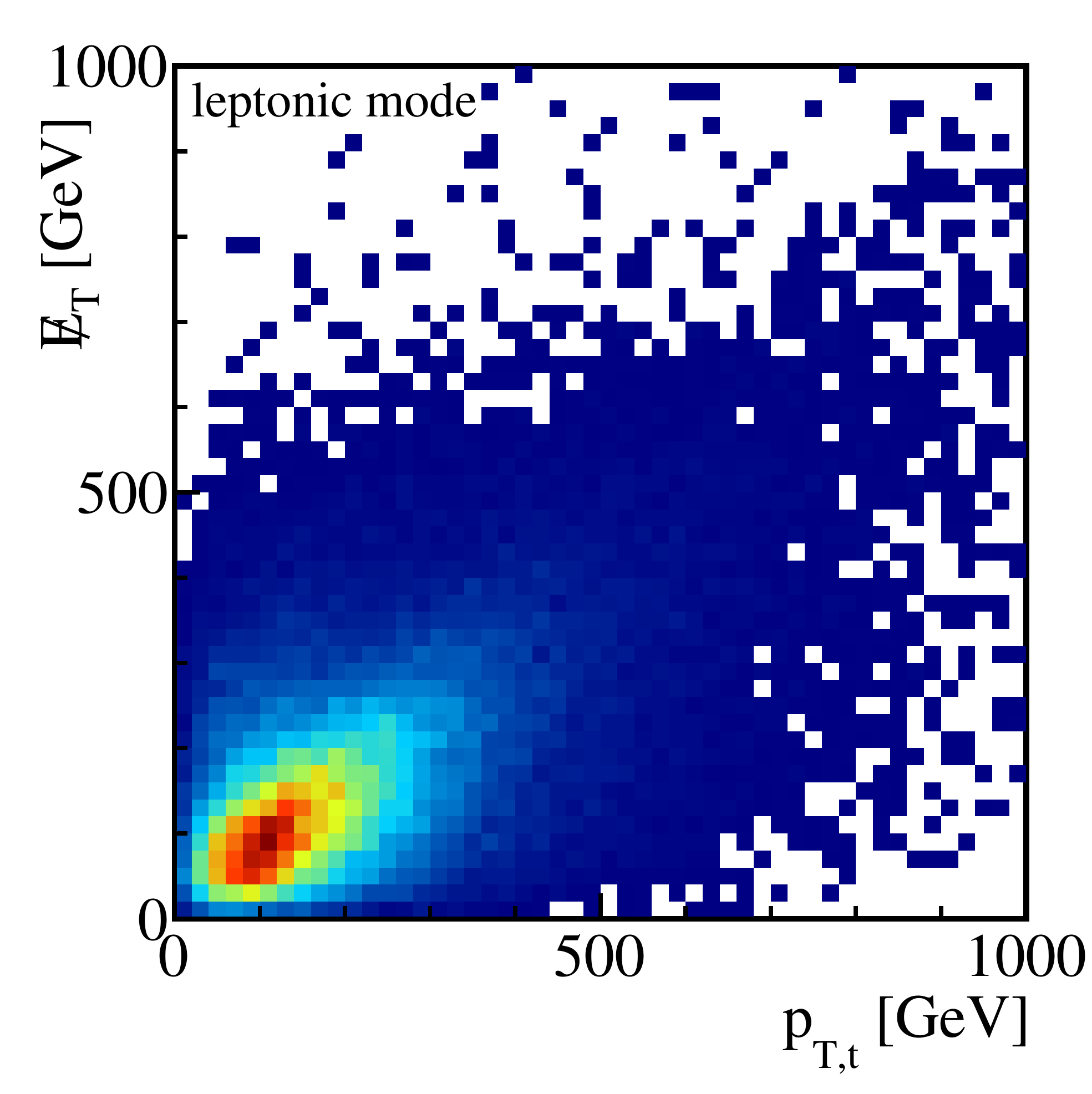}     
    \caption{2D 
    $(p_{T,t}, \slashed{E}_T)$ distribution for the hadronic (top) and leptonic (bottom) channels.  
    The dense regions are shown in red.
    Notice that the leptonic final state provides a softer $\slashed{E}_T$ profile as
    the neutrino momentum from the top-quark decay partly cancels the missing transverse energy generated by neutralinos.
    }
    \label{fig:2d}
    \end{center}
\end{figure}

We start our analysis by vetoing isolated leptons with $p_{T \ell}>10$ GeV, $|\eta_\ell|<2.5$ and requiring large missing 
energy: $\slashed{E}_T>300$~GeV. For the jet reconstruction, we have used the calorimeter tower information obtained by \textsc{Delphes3}.
Only the cells with the transverse energy larger than 0.5 GeV are taken into account. 
We take advantage of the hadronic top in the boosted regime by reclustering the calorimeter towers into a fat-jet with the radius parameter of $R=1.5$ using the Cambridge/Aachen jet algorithm implemented in \textsc{Fastjet}~\cite{fastjet2}. 
We require at least one fat-jet with $p_{TJ}>200$~GeV, $|\eta_{J}|<2.5$
and this jet must be top-tagged by the \textsc{HepTopTagger}~\cite{Plehn:2009rk,Plehn:2010st}. 
The \textsc{HepTopTagger} was initially designed to reconstruct mildly boosted top-quarks with $p_{T,t}\sim m_t$.
However, large flexibility of the algorithm allows to achieve a good tagging efficiency $\sim 30\%$ for highly boosted tops, $p_{T,t} \gtrsim 400$~GeV, keeping the fake rate within the level of $3\%$. 
The red solid histogram in Fig.~\ref{fig:toptag} shows the top-tagging efficiency as a function of the top-quark $p_{T,t}$.
Also shown in Fig.~\ref{fig:toptag} by the blue dotted histogram is the mistag rate for QCD jets as a function of the fat-jet $p_{T,J}$.
The tagging efficiency is estimated in the signal sample, while the mistag rate is obtained in the $Z$+jets sample.

To further suppress the $Z$+jets background, we also require at least one of the three subjets -- inside the fat-jet -- to be $b$-tagged,
assuming the $b$-tagging efficiency of 70\% and the mistag rate of 1\%. After a successful top tagging, we remove the tagged top-jet constituents 
and recluster the remaining calorimeter towers, but now with the anti-k$_{T}$ jet algorithm with $R=0.4$, $p_{Tj}>30$~GeV and $|\eta_j|<2.5$. To suppress the dominant 
$t \bar{t}$ background, we apply an extra-jet veto, $n_j=0$. 
We have checked that relaxing this condition (e.g. $n_j \le 1$) only deteriorates the sensitivity due to 
the overwhelming contribution from the $t \bar t$ background, even when rejecting extra $b$-tagged jets.

\begin{figure}[t!]
    \begin{center}
          \includegraphics[scale=0.35]{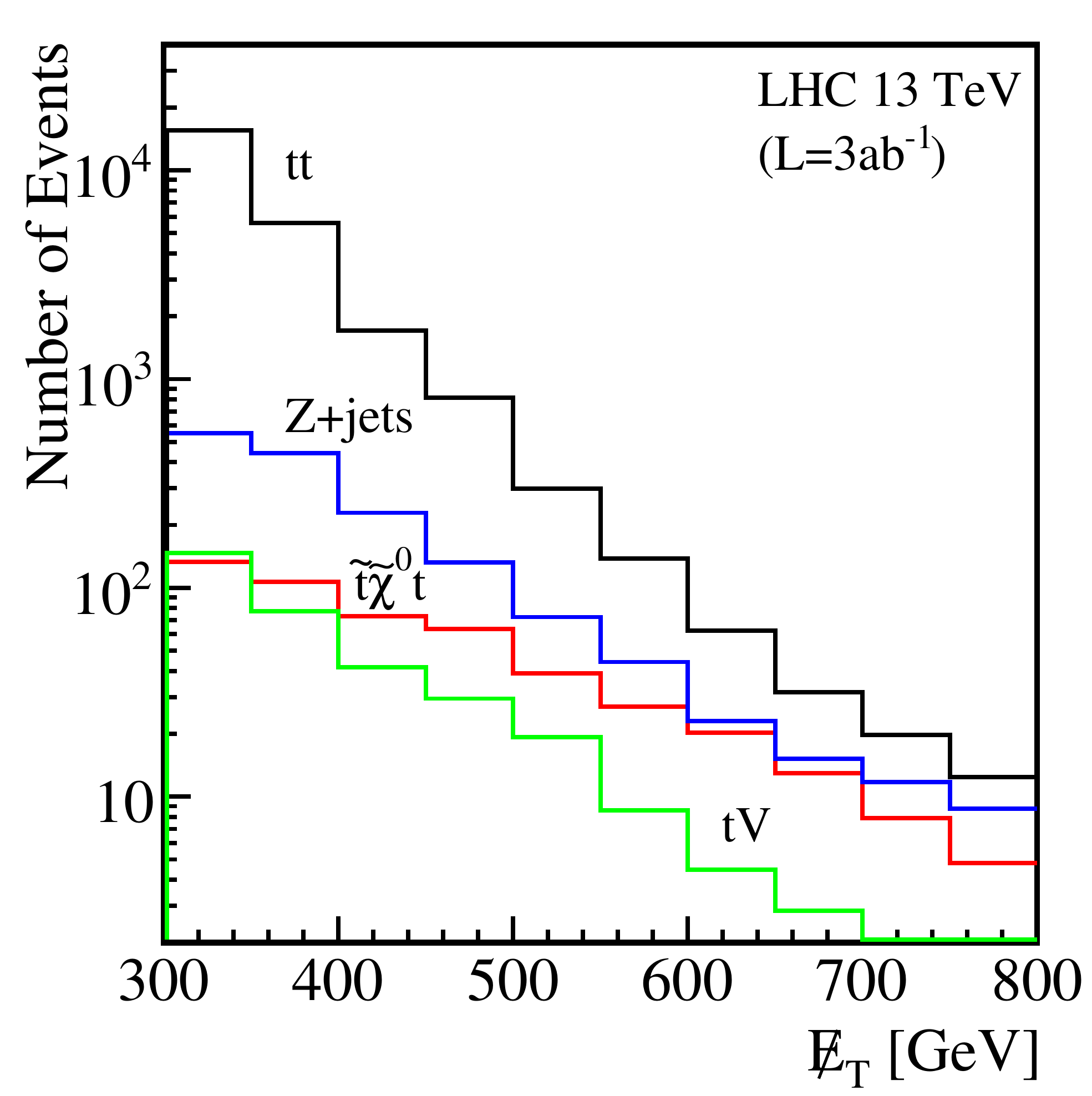}          
    \caption{Signal  and backgrounds transverse missing energy $\slashed{E}_T$ distributions for the at the 13~TeV LHC. 
    We consider the signal benchmark point  $m_{\tilde{t}_1}=342$~GeV, ${m_{\chi^0_i}=334}$~GeV.}
    \label{fig:met}
    \end{center}
\end{figure}

It is interesting to compare the hadronic and leptonic mono-top channels. 
In Fig.~\ref{fig:concept} we show the $\slashed{E}_T$ distributions for the hadronic (red) and leptonic (blue)
channels of the $pp \to t \tilde t_1 \tilde \chi_{1(2)}^0$ process in the solid lines. 
The corresponding dotted histograms are the truth level top-quark $p_{T,t}$ 
distributions. 
The hadronic final state has a much larger rate 
due to the greater hadronic branching ratio, $\mathcal{BR}_{\rm had} \sim 0.68$, of the top-quark.
We can also see that the hadronic channel leads to more energetic $\slashed{E}_T$ distribution 
in comparison with the leptonic one, which are shown in the corresponding solid lines.
The source of this larger $\slashed{E}_T$ can be appreciated by looking at
the 2D $(p_{T,t}, \slashed{E}_T)$ distributions
shown in the top (for the hadronic channel) and bottom (for the leptonic channel) plots in Fig.~\ref{fig:2d}.
While the hadronic top fully balances the transverse momentum with the two neutralinos in the final state ($\overrightarrow{\slashed{p}}_T = - \overrightarrow{p}_{t,T}$), in the leptonic channel the $\slashed{E}_T$ generated by the neutralinos is partly cancelled by the neutrino from the top-quark decay. 
It is also worth noting that this cancellation in $\slashed{E}_T$ in the leptonic channel is more significant 
for larger $\slashed{E}_T$ bins, where the hadronic top-tagging becomes most efficient 
due to the boosted kinematics of the top-quark as explicitly seen in the tagged top $p_{T,t}$ distribution (red dashed).  
As a result, the number of events with $\slashed{E}_T \gtrsim 400$~GeV in the hadronic channel exceeds that in the leptonic channel
even after taking the top-tagging efficiency into account. 

In Fig.~\ref{fig:met} we display the $\slashed{E}_T$ distribution for the signal and background samples after the full event selections. 
We observe that the signal to background ratio, $\mathcal{S}/\mathcal{B}$, increases in the region with large $\slashed{E}_T$.  
To exploit this feature
we divide our analysis into three signal regions, $\mathcal{SR}$, that differ by the $\slashed{E}_T$ requirement as
$\slashed{E}_T > 400, 500$ and $600$~GeV. The full cut-flow analysis is provided in Tab.~\ref{tab:cuts}. \medskip

\begin{table*}[!t]
\centering
\begin{tabular}{l || rrr  || r | r  | r | r || r}
  \multicolumn{1}{c||}{} &
  \multicolumn{3}{c||}{$t \tilde{t}_1\tilde{\chi}_{1(2)}$}&
  \multicolumn{1}{c|}{$\bar{t}t$} &
    \multicolumn{1}{c|}{$tW$} &
  \multicolumn{1}{c|}{$tZ$} &
      \multicolumn{1}{c||}{$Z$+jets} &
      \multicolumn{1}{c}{Total BG} \\
  \hline
    \multicolumn{1}{c||}{model point} &(342  334) &(394  368)&(394  386)&&&&&\\
    \hline
$n_\ell = 0$, top-tag, $p_{TJ}>200$~GeV, $\slashed{E}_T>300$~GeV 
&2103 &1275.8&1245.5 &128924 &8821&1260&68923 &207928 \cr
 $b$-tag in $t_{\rm tag}$ (70\%, 1\%$\times 3$ combinatorial)
&1472 &893.0&871.8&90246 &6174&882 &2068 &99370 \\
 $n_j = 0$ ($p_{Tj}>30$~GeV,  $|\eta_j|<2.5$) &
 507.1&240.9&288.4 &24248 &2520 &168&1550 & 28486  \cr
\hline
$\mathcal{SR}1$: $\slashed{E}_T>400$~GeV  &267.0 &124.4&160.8&3114&504&52.5&556.5&4227 \cr
$\mathcal{SR}2$: $\slashed{E}_T>500$~GeV  &130.4&57.8&83.5 &595.6&105&25.2&195.2 &921.0 \cr
$\mathcal{SR}3$: $\slashed{E}_T>600$~GeV  &64.5&26.5&44.7&151.5&29.4&10.5&74.7 &266.1 \cr
\hline
$\mathcal{S}/\mathcal{B}$ and $\mathcal{S}/\sqrt{\mathcal{B}}$ for $\mathcal{SR}1$ & (0.06,4.1) & (0.03,1.9)& (0.04,2.5) \cr
\ \ \ \ \ \ \ \ \ \ \ \ \ \ \ \ \ \ \ \ \, for $\mathcal{SR}2$ & (0.14,4.3) & (0.06,1.9)& (0.09,2.8) \cr
\ \ \ \ \ \ \ \ \ \ \ \ \ \ \ \ \ \ \ \ \, for $\mathcal{SR}3$ & (0.24,4.0) & (0.1,1.6)& (0.17,2.7) \cr
\hline
\end{tabular} 
\caption{Cut-flow analysis for the signal and backgrounds at the LHC $\sqrt{s}=13$~TeV. 
The number of signal and background events are shown assuming $\mathcal{L}=3$~ab$^{-1}$.}
\label{tab:cuts}
\end{table*}

\section{Results}
\label{sec:results}

\begin{figure*}[t!]
\centering
         \includegraphics[scale=0.45]{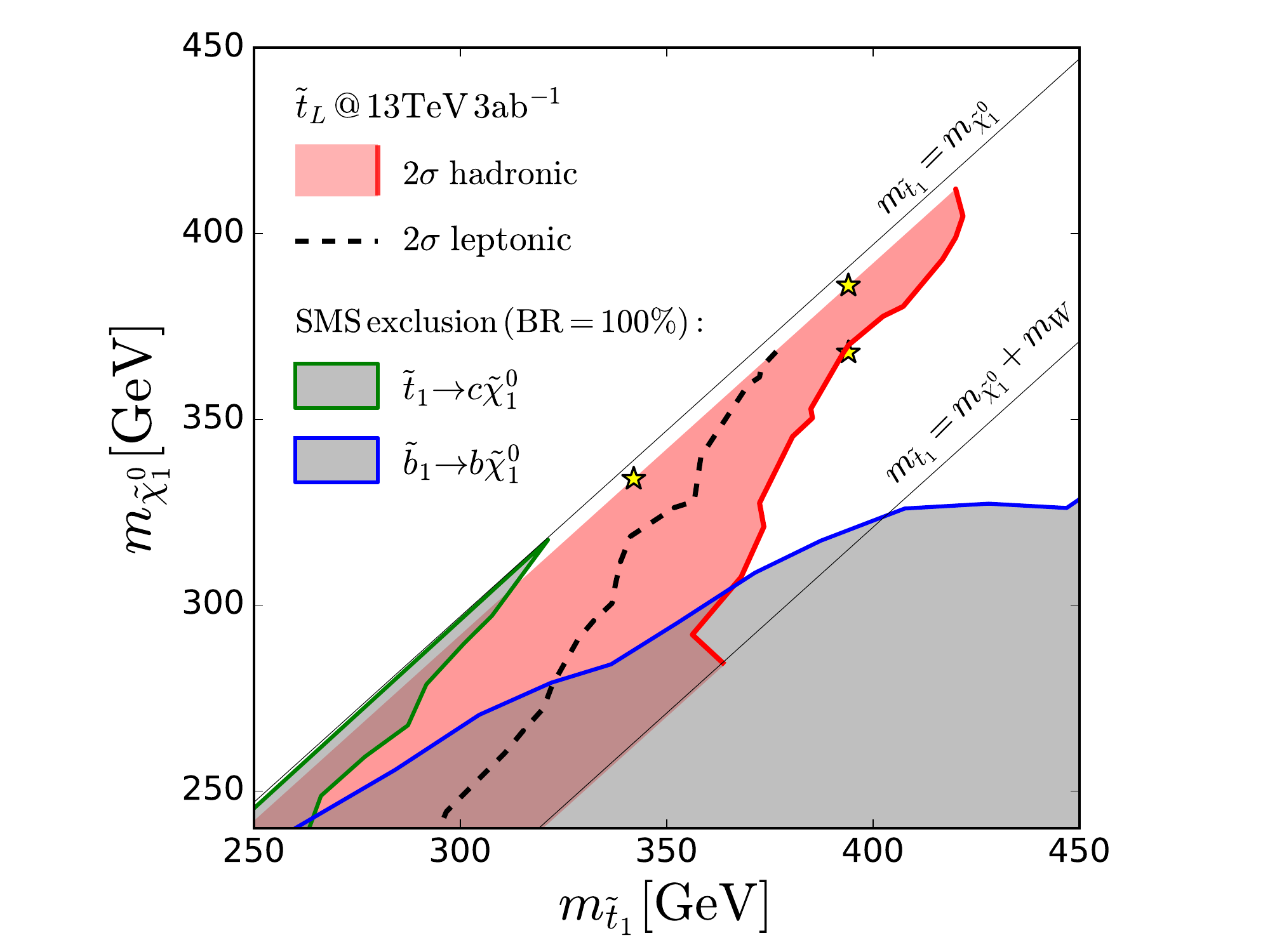}
          \hspace{-1.2cm}          
          \includegraphics[scale=0.45]{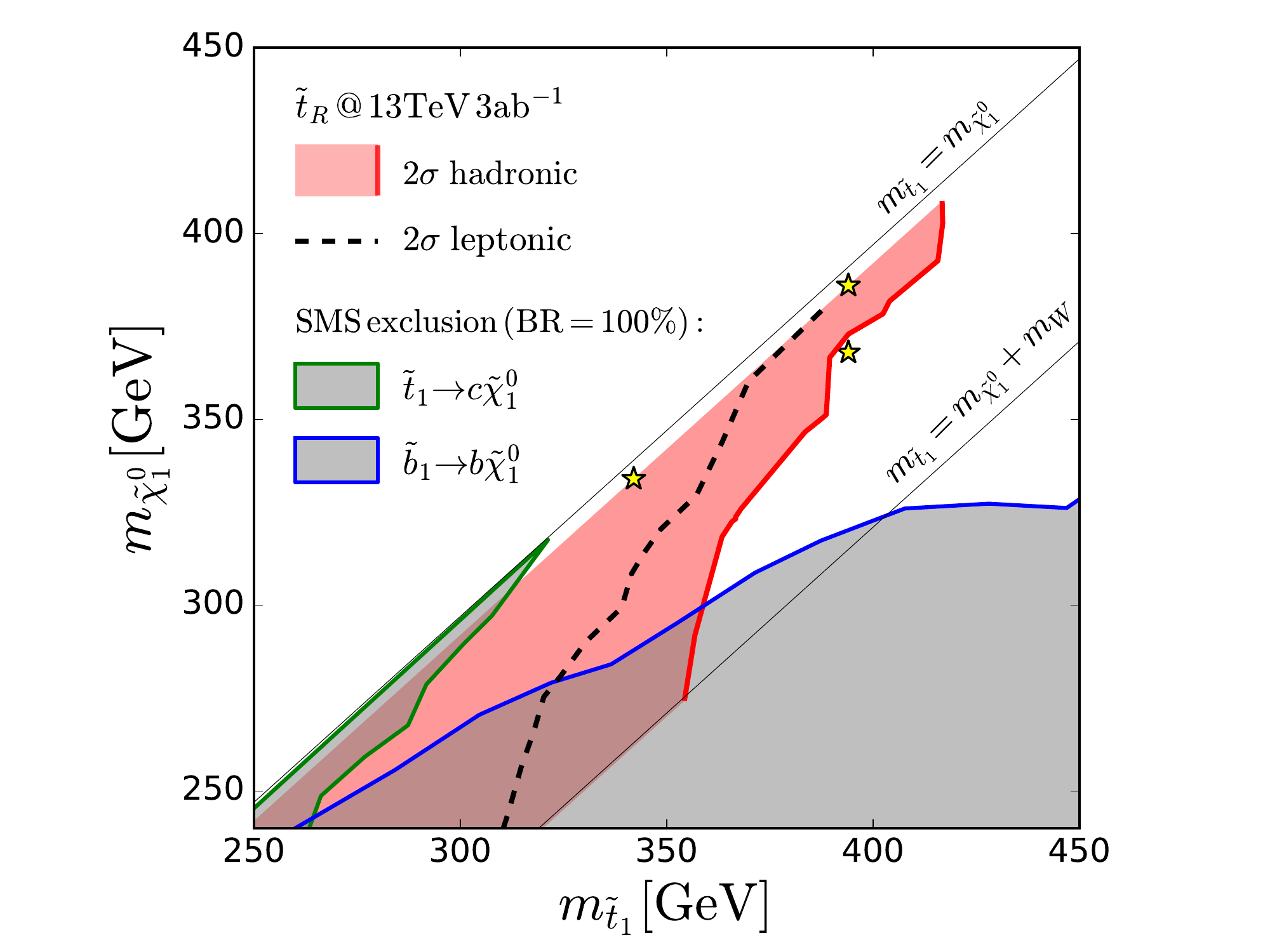}          
    \caption{
Expected 95\% CL sensitivities at the 13~TeV high luminosity LHC $\mathcal{L} = 3$~ab$^{-1}$.
The light red regions in the left and right panels correspond to the 2\,$\sigma$ regions for the $\tilde t_1 = \tilde t_L$ and $\tilde t_1 = \tilde t_R$ cases, respectively, obtained by the hadronic mono-top analysis presented in this paper.
The black dashed contours are the 2\,$\sigma$ regions obtained by the leptonic mono-top analysis shown in \cite{Goncalves:2016tft}.  
The current 95\% CL excluded regions based on Simplified Models for SUSY (SMS) assuming 100\% BR are also shown in grey. 
The  blue and green curves are obtained from the di-$b$-jet \cite{ATLAS-CONF-2015-066} and mono-jet \cite{Aaboud:2016tnv} analyses
based on the 13~TeV data with $\mathcal{L}= 3.2$ fb$^{-1}$.
}
    \label{fig:sensitivity}
\end{figure*}

We now show the performance of our hadronic mono-top analysis 
assuming the 13~TeV LHC with ${\mathcal{L}=3}$~ab$^{-1}$
and compare it with the leptonic analysis studied in \cite{Goncalves:2016tft}.
In the left ($\tilde t_1 = \tilde t_L$) and the right ($\tilde t_1 = \tilde t_R$) panels of Fig.~\ref{fig:sensitivity},
the 95\% CL sensitivity regions derived from the hadronic mono-top analysis are highlighted by the light red colour.
When deriving the sensitivity, we choose the most sensitive signal region with the largest $\mathcal{S}/\sqrt{\mathcal{B}}$.
To ensure that the systematic uncertainty is under control, we only consider
the regions with $\mathcal{S}/\mathcal{B}>0.1$. The three benchmark points in Tab.~\ref{tab:cuts} are denoted by the stars.
As can be seen, the performance of this analysis is not sensitive to whether the $\tilde t_1$ is dominantly $\tilde t_L$ or $\tilde t_R$.  
In comparison, we also show the 95\% CL sensitivity derived from the leptonic mono-top analysis \cite{Goncalves:2016tft}
with the black dashed curve.
It is clear that the sensitivity from the hadronic analysis is superior in all regions.
For example, in the most compressed ($m_{\tilde t_1} \simeq m_{\tilde \chi_1^0}$) region,
the sensitivity reaches $m_{\tilde t_1} \sim 420$ GeV for the hadronic channel,
whilst the sensitivity for the leptonic channel is limited up to $m_{\tilde t_1} \sim 380$ GeV
for both $\tilde t_L$ and $\tilde t_R$ cases.
As we have discussed in detail in the previous section, the superiority of the hadronic channel
is attributed to $\mathcal{BR}_{\rm had} \gg \mathcal{BR}_{\rm lep}$ and 
the absence of the partial cancellation in the $\slashed{E}_T$ between the neutralinos and the neutrino in the leptonic channel.
We also superimpose the current exclusion limits\footnote{The preliminary result of CMS \cite{CMS:2016hxa}
claims their excluded region reaches $m_{\tilde t_1} \sim 380$ GeV in the most mass degenerate region
assuming $\mathcal{BR}(\tilde t_1 \to b f \bar f^\prime \tilde \chi_1^0) = 100\%$.
This strong exclusion is achieved by explicitly looking at soft $b$-jets from $\tilde t_1 \to b f \bar f^\prime \tilde \chi_1^0$.
This technology would also improve the sensitivity of our mono-top analysis to the $pp \to \tilde t_1 t \tilde \chi_{1(2)}^0$ process. We however leave this analysis for future work.} 
derived in 
simplified models assuming 100\% branching ratios of the $\tilde t_1$.
The grey region surrounded by the green curve is excluded by the 13 TeV ATLAS mono-jet analysis \cite{Aaboud:2016tnv} 
assuming $\mathcal{BR}(\tilde t_1 \to c \tilde \chi_1^0) = 100\%$,
whereas the grey region with blue curve is excluded by 
the 13 TeV ATLAS di-$b$ jet analysis \cite{ATLAS-CONF-2015-066}
assuming $\mathcal{BR}(\tilde b_1 \to b \tilde \chi_1^0) = 100\%$.
Strictly speaking, the latter limit cannot be directly applied to the $(m_{\tilde t_1}, m_{\tilde \chi_1^0})$ plane.
However, in our set up with the $\tilde t_1$ predominantly decaying into $b$ and higgsino-like $\tilde \chi_1^\pm$
with $m_{\tilde \chi_1^\pm} \simeq m_{\tilde \chi_1^0}$,
both production rates and event topologies are similar between $pp \to \tilde t_1 \tilde t_1 \to b \tilde \chi_1^+ b \tilde \chi_1^-$ and $pp \to \tilde b_1 \tilde b_1 \to b \tilde \chi_1^0 b \tilde \chi_1^0$. So, this limit can be applied at least approximately.
We also comment that these published exclusion limits are sensitive to the $\tilde t_1$ decay.
On the other hand, the mono-top analysis presented in this paper is less sensitive to it 
since the high $p_T$ objects used in the analysis 
are not originated from the $\tilde t_1$ decay but from the top-quark decay.

\section{Conclusion}
\label{sec:conclusion}

We have studied a class of Natural SUSY models where 
the stop and higgsinos have almost equal masses.
It has been known that this compressed region can be 
most effectively searched for by the mono-jet channel, 
exploiting hard QCD initial state radiation.
The drawback of the mono-jet channel is that
the hight $p_T$ jet is entirely controlled 
by QCD and does not carry  information on
the stop and higgsino sectors.
Indeed, finding this signature does not necessarily indicate 
the existence of the light stop and higgsino. 
In order to probe the stop and higgsino sectors,
another channel providing orthogonal information is required.

In this paper we have studied a supersymmetric version of the $t \bar t H$ process, 
namely $t \tilde{t} \tilde{\chi}^0_{1(2)}$ production.
In the region where the mass spectrum is compressed ($m_{\tilde t_1} \simeq m_{\tilde \chi_1^0}$)
this process leads to a distinctive mono-top signature.
The three particle production process $pp \to t \tilde{t} \tilde{\chi}^0_{1(2)}$ can have observably large rates
only if both the stop and higgsinos are significantly light. 
The mono-top signature can thus be regarded as the smoking gun signature 
of the compressed region of the Natural SUSY scenario.

We focused in this article on the hadronic final state of the mono-top signature
with an obvious advantage of $\mathcal{BR}_{\rm had} \gg \mathcal{BR}_{\rm lep}$.
In order to discriminate the signal from backgrounds,
we have used \textsc{HepTopTagger} to ``tag'' a boosted hadronic top in the signal. 
We found a superior performance in the sensitivity for the hadronic mono-top analysis
over the previously studied leptonic analysis \cite{Goncalves:2016tft}.
This is attributed not only to ${\mathcal{BR}_{\rm had} \gg \mathcal{BR}_{\rm lep}}$
but also to the fact that $\slashed{E}_T$ is harder in the hadronic channel than in the leptonic one 
because the $\slashed{E}_T$ generated by the neutralinos is partially cancelled by the neutrino 
from the top-quark decay in the leptonic channel.
After performing MC simulation including the detector effects, we have found 
the sensitivity in the hadronic mono-top analysis reaches ${m_{\tilde t_1} \simeq 420}$~GeV,
exhibiting a significant improvement over the leptonic analysis whose reach is $m_{\tilde t_1} \simeq 380$~GeV.
We also observed that in order to suppress the background 
very large $\slashed{E}_T$ (e.g.~$\slashed{E}_T > 400-600$ GeV) must be required.  
The $\slashed{E}_T$ is highly correlated to the top-quark $p_T$
and the top-tagging becomes most efficient in the hight $p_T$ region.
We therefore expect that the hadronic mono-top channel works also well for the light stop and higgsino 
searches  at future 100~TeV $pp$ colliders.


\section*{Acknowledgements}

DG and KS were supported by STFC through the IPPP grant. The work of DG was also partly funded by the U.S. 
National Science Foundation under grant PHY-1519175. The work of KS is  partially supported by the National 
Science Centre, Poland, under research grants DEC-2014/15/B/ST2/02157  and DEC-2015/18/M/ST2/00054.
MT is supported by World Premier  International Research Center Initiative (WPI Initiative), MEXT, Japan.
MT is supported by Grant-in-Aid for Scientific Research Numbers JP16H03991 and JP16H02176. 
DG and MT are grateful to the Mainz Institute for Theoretical Physics (MITP) for its hospitality and its partial support 
in the early stages of this work. 

\newpage

\bibliographystyle{apsrev}
\bibliography{draft}

\end{document}